\documentclass[aps,prb,twocolumn,groupedaddress,showpacs,amsmath,amssymb]{revtex4}

\usepackage[english]{babel}

\usepackage{graphicx}
\usepackage{hyperref}
\usepackage{amsmath,amsthm,amssymb}

\DeclareMathOperator{\Var}{Var}
\DeclareMathOperator{\Cov}{Cov}
\DeclareMathOperator{\FWHM}{FWHM}

\begin{document}

\raggedbottom

\title{Probing atomic environments in alloys by electron spectroscopy}
\author{T. L. Underwood}
\author{G. J. Ackland}
\author{R. J. Cole}
        
\affiliation{School of Physics and Astronomy, SUPA, The University of Edinburgh, Edinburgh,
EH9 3JZ, UK }

\date{\today}


\begin{abstract}
In alloys exhibiting substitutional disorder, the variety of atomic environments manifests itself as a
`disorder broadening' in their core level binding energy spectra. Disorder broadening can be 
measured experimentally, and in principle can be used to deduce information about specific atomic environments 
within a sample. However, progress in this endeavor is hampered by the lack of a model for this
phenomenon which can treat complex systems.
In this work we describe such a model. The model is used to elucidate the relationship between charge transfer,
atomic environment, and disorder broadening in complex systems, with a focus on the problem of characterizing 
the interface quality of CuNi multilayers. We also validate the model against the results of \emph{ab initio}
electronic structure calculations. Several counterintuitive aspects of the disorder broadening phenomenon are uncovered, an 
understanding of which is essential for the correct interpretation of experimental results. For instance, it
is shown that systems with inhomogeneous concentration profiles can exhibit disorder broadenings significantly
larger than random alloys. Furthermore in some systems a `disorder narrowing' is even possible.
\end{abstract}

\maketitle

\section{Introduction}\label{sec:intro}
An atom's core level binding energies depend on its chemical species. This fact has been exploited for decades in order to determine the proportion
of different species within a given sample via core level spectroscopy - which provides the distribution of core electron binding energies within a
sample. An atom's binding energies also depend on its environment, i.e., the species of its surrounding atoms. 
For instance, in a Cu metal all atoms have the same environment, and hence the binding energies for a given Cu core level type, e.g. 2p$_{3/2}$, will be 
the same for all atoms. By contrast, in a CuPd alloy exhibiting substitutional disorder the Cu atoms exhibit a variety of environments, and hence
also a variety of 2p$_{3/2}$ binding energies. Such a dispersion in the binding energies has been observed in many alloy systems
\cite{Cole_1997_PRL,Cole_1998,Lewis_1999,Newton_2000,Newton_2004,Marten_2009,Holmstrom_2006,Granroth_2009}, 
and in theory can be used to deduce information about specific environments within a sample, including the concentration profile on
the atomic scale. The viability of achieving this via high kinetic energy photoelectron spectroscopy (HIKE) has recently been demonstrated
\cite{Holmstrom_2006,Granroth_2009}. This is promising because HIKE, unlike other widely-used techniques, is both bulk-sensitive and 
non-destructive. Such environment-resolved spectroscopy would prove useful to the many research areas involving alloys in which segregation plays
a crucial role, e.g. metal embrittlement \cite{Duscher_2004,Schweinfest_2004,Yamaguchi_2005,Chen_2010,Luo_2011,Kang_2013}, and nanocatalyst
design \cite{vanderVliet_2012,Chunhua_2013}.
However, success in this endeavor hinges upon a solid understanding of the relationship between an atom's environment, its 
electronic structure, and its core level binding energies in alloys.

The central quantity with regards to environment-resolved spectroscopy in alloys is the (alloy-metal) core level shift 
(CLS), which for a core level bound to an $X$ site $i$ is defined as
\begin{equation}\label{CLS_def}
\Delta E^{\text{B}}_i=E^{\text{B}}_i-E^{\text{B}}_{X\text{met}},
\end{equation}
where $E^{\text{B}}_i$ is the binding energy of the core level, and $E^{\text{B}}_{X\text{met}}$ is the binding energy of the core level belonging
to the type under consideration in a pure $X$ metal. Note that this is a site-dependent quantity; one must determine $\Delta E^{\text{B}}_i$
for all $X$ sites in the system under consideration in order to determine the $X$ CLS distribution. This is problematic for systems 
exhibiting substitutional disorder on account of their lack of periodicity - which is a prerequisite for treatment within the conventional theoretical 
framework exploited by most \emph{ab initio} methods. One way around this problem is to approximate the system under consideration as periodic, 
but with a large unit cell, i.e., a supercell. In this \emph{supercell approximation} one determines $\Delta E^{\text{B}}_i$ for all $X$ sites, and hopes 
that the range of environments exhibited by these sites is representative of the `true' (non-periodic) system, and hence will 
result in an accurate representation of the true $X$ CLS distribution. 

Random alloys - the archetype of disordered alloys in which there are no correlations between the species of sites - are the most tractable
system exhibiting substitutional disorder to treat theoretically, and have been the focus of both experimental
\cite{Cole_1997_PRL,Cole_1998,Lewis_1999,Newton_2000,Newton_2004,Marten_2009,Olovsson_2011} and theoretical 
\cite{Cole_1997,Cole_1997_PRL,Cole_1998,Underwood_2009,Underwood_2010,Underwood_2013,Faulkner_1998_PRL,Marten_2005,Marten_2009,Olovsson_2011}
attempts to understand the distribution of CLSs in alloys. In this context the dispersion of $X$ CLSs
is known as `disorder broadening' on account of the increased width of, for example, the Cu CLS distribution for a CuPd random alloy relative to that 
for a Cu metal.
Sophisticated \emph{ab initio} models utilizing the supercell approximation have provided insight into disorder broadening in these systems
\cite{Faulkner_1998_PRL,Marten_2005,Marten_2009,Olovsson_2011}. However, most systems of practical interest cannot 
be idealized as random alloys, and unfortunately the complexity of these systems is such that a description of their 
disorder broadenings using \emph{ab initio} models is intractable within the supercell approximation.

Accurate methods do exist which do not resort to the supercell approximation. 
\emph{Ab initio} methods rooted in the coherent potential approximation (CPA)\cite{Soven_1967,Johnson_1986,Johnson_1990} have been shown to provide
excellent agreement with experiment \cite{Abrikosov_2001,Olovsson_2002,Olovsson_2005_fcc,Olovsson_2005_surface,Olovsson_2011}. Furthermore, they
can treat complex systems \cite{Olovsson_2005_surface,Holmstrom_2006,Olovsson_2011}.
However, CPA-based approaches cannot provide detailed information regarding the disorder broadening in complex systems - nor even random alloys.
While phenomenological models have been developed which can provide such information 
\cite{Cole_1997,Cole_1997_PRL,Cole_1998,Underwood_2009,Underwood_2013}, their accuracy has been questioned 
\cite{Weightman_1998_PRL_comment,Faulkner_1998_PRL_reply,Methfessel_2000}. One criticism is that these models do not take into account `final state 
effects' associated with changes in the valence electron density after photoemission. It has even been claimed that the complexity of the relationship 
between CLSs and environment in alloys precludes an accurate alternative to \emph{ab initio} methods \cite{Methfessel_2000}.

\begin{figure}
\centering
\includegraphics[width=0.5\textwidth]{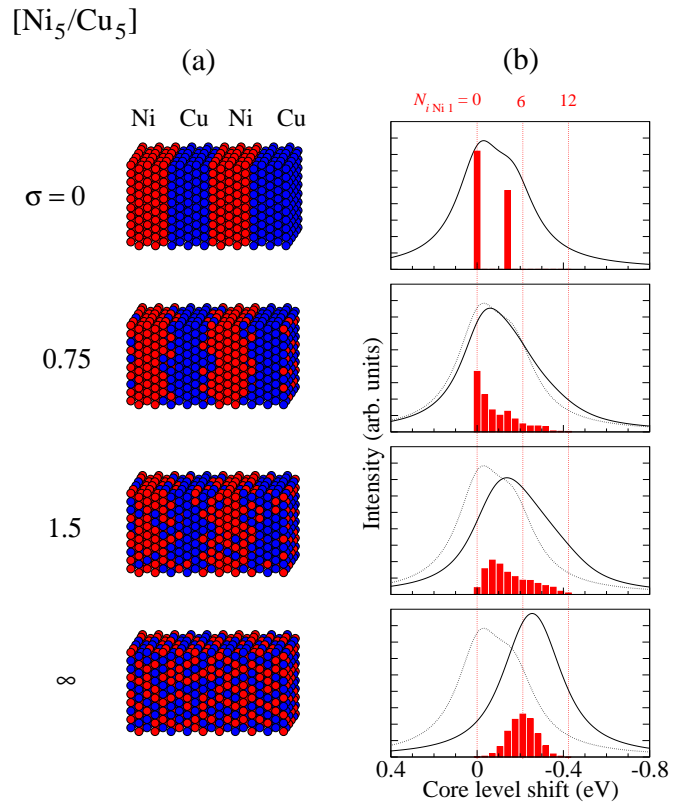}
\caption
{(Color online) Schematic illustration of [Ni$_5$/Cu$_5$] at various $\sigma$ (a), and the
corresponding model spectra (b). In (b), each red bar corresponds to a different value of $N_{i\text{Ni}1}$:
the ordinate of the bar is the CLS corresponding to $N_{i\text{Ni}1}$ according to Eqn. \eqref{DeltaEBi_nn_bin}, and its height
reflects the frequency of Cu sites with that value of $N_{i\text{Ni}1}$. The bars corresponding to $N_{i\text{Ni}1}=0$, 6, and 12 are
indicated. The dotted curves in (b) are spectra for $\sigma=0$.}
\label{fig:figure_Cu5Ni5}
\end{figure}

Here we present an accurate phenomenological model for CLSs in alloys which relates $\Delta E^{\text{B}}_i$ - including the final state contribution - to the 
environment of site $i$. The model, like previous approaches, is charge-transfer based, and provides a simple framework for rationalizing the disorder 
broadening phenomenon. The layout of this work is as follows. In Sec. \ref{sec:model} we review the theory which underpins the model, and
derive expressions for CLSs which apply to a wide range of alloy systems. We then apply the model to the
problem of characterizing the interface quality of metallic multilayers. This problem has received significant attention on account of its 
importance to nanotechnology \cite{Holmstrom_2004,Holmstrom_2006,Granroth_2009,Olovsson_2011}, and can be restated as follows: what is the degree of 
`interface roughening' $\sigma$ in a given sample? Fig. \ref{fig:figure_Cu5Ni5}(a) provides an illustration of the multilayer system [Ni$_5$/Cu$_5$] with 
various $\sigma$ - where the square brackets signify that the system at $\sigma=0$ consists of a 10 monolayer stack Ni$_5$/Cu$_5$ repeated throughout all 
space. In this regard, Sec. \ref{sec:comp_details} contains details of our calculations, and our results are presented in Sec. \ref{sec:results}.
Finally, in Sec. \ref{sec:summary} we summarize our main findings, and discuss the limitations of the model, and intentions for future work.
Note that throughout this work we use Hartree atomic units unless otherwise stated. To transform energies within Hartree atomic units to eV the former 
quantity should be multiplied by a factor 27.2114.

\section{Theory}\label{sec:model}
Our model is based upon the charge-excess functional model \cite{Bruno_2003} within the non-random approximation \cite{Drchal_2006,Underwood_2013} 
(NRA-CEFM). After briefly reviewing the NRA-CEFM in Sec. \ref{sec:NRACEFM_review}, in Sec. \ref{sec:NRACEFM_DeltaEBi} we use it to derive 
a relation between $\Delta E^{\text{B}}_i$ and the environment of site $i$. In the subsequent subsections this relation is itself used to derive expressions 
for the mean and full width at half maximum (FWHM) of various CLS distributions. While our expression for $\Delta E^{\text{B}}_i$ presented in 
Sec.  \ref{sec:NRACEFM_DeltaEBi} is valid for \emph{any} alloy - subject to the assumptions which underpin the NRA-CEFM - our expressions for means and 
FWHMs apply only to the class of alloys which can be described as having a concentration profile which varies along one direction from monolayer to 
monolayer. To elaborate, for an alloy belonging to this class, the quantities $c_X^l$ for all $X$ and $l$ describe the concentration profile, where 
$c_X^l$ denotes the concentration of $X$ sites within monolayer $l$, and we are assuming that the species of sites belonging to a given monolayer $l$ are 
assigned randomly in the appropriate concentrations. Note that this class includes random alloys, for which $c_X^l=c_X$ for all $l$, where $c_X$ denotes 
the global concentration of species $X$. Furthermore, note that we use the following convention for labeling the monolayers: monolayer $l$ is the $l$th
monolayer in the direction in which the concentrations of each species are varying.

\subsection{The charge-excess functional model in the non-random approximation}\label{sec:NRACEFM_review}
The charge-excess functional model (CEFM) \cite{Bruno_2003} has been shown to provide an accurate description of
the charge distribution in disordered alloys \cite{Bruno_2003,Drchal_2006,Bruno_2008}. It has also been shown to
provide an accurate description of energy differences between alloy configurations with the same
\emph{composition} \cite{Bruno_2008} - where by composition we mean a specification of the underlying lattice and 
the quantities $c_X$ for all $X$. In the CEFM the 
alloy energy is postulated to take the form
\begin{equation}\label{E_gen_def}
E=E_0+\frac{1}{2}\sum_ia_i(Q_i-b_i)^2+E_M,
\end{equation}
where
\begin{equation}\label{EM_def}
E_M=\frac{1}{2}\sum_iQ_iV_i
\end{equation}
is the Madelung energy, $V_i$ is the Madelung potential of site $i$, $Q_i$ is the net charge on site $i$, $a_i$ is the strength of the `local
interactions' within site $i$ which act to keep the charge of site $i$ at its `bare' value $b_i$, and $E_0$ is a constant.
For all $X$ sites, $a_i$ and $b_i$ take the same values $a_X$ and $b_X$ respectively.
Minimizing $E$ subject to the constraint of charge neutrality leads to the following expression \cite{Bruno_2003}:
\begin{equation}\label{QV_rel}
V_i=-a_iQ_i+k_i,
\end{equation}
where
\begin{equation}\label{k_def}
k_i=a_ib_i+\lambda,
\end{equation}
and $\lambda$ is a Lagrange multiplier added to enforce charge neutrality. Eqn. \eqref{QV_rel} describes the
\emph{Q-V} relations, which are borne out in \emph{ab initio} calculations to a high degree of accuracy 
\cite{Faulkner_1995,Faulkner_1997,Ruban_2002,Bruno_2008}. This partly explains the success of the CEFM.

Eqn. \eqref{E_gen_def} can be derived within the class of generalized coherent potential approximations (GCPAs) described
by Bruno \emph{et al.} \cite{Bruno_2008}, with the additional - essentially exact \cite{Bruno_2008} - assumption that charge 
transfer due to Madelung interactions 
\footnote{In this work the term `Madelung interactions' is used as a synonym for `inter-site Coulomb interactions'.}
is small. Here we consider the CEFM within this framework, in which case $b_X$ is equivalent to the 
charge of an $X$ site embedded in the GCPA effective medium for the system under consideration, given the constraint that 
the site's Madelung potential is 0. 
In other words, $b_X$ is the charge of an $X$ site embedded in the effective medium if the Madelung interactions 
are `switched off'. Furthermore, $a_X$ is the linear response coefficient relating the Madelung potential of the $X$ site to its
perturbation from $b_X$, and $E_0$ is the energy of the alloy if the Madelung interactions between sites are switched off - which
can be expressed as
\begin{equation}\label{E0_def}
E_0=\mathcal{N}\sum_Xc_X\mathcal{E}_X,
\end{equation}
where $\mathcal{E}_X$ is the energy of an $X$ site embedded in the effective medium if Madelung interactions are switched off, and
$\mathcal{N}$ is the total number of sites in the system.
Conventional CPA calculations do not take into account Madelung interactions, and hence for GCPA theories based upon CPA effective media 
$b_X$ is simply the charge of an $X$ site obtained from a conventional CPA calculation, and $\mathcal{E}_X$ is the corresponding $X$ energy.
Calculations reveal that GCPA effective media are numerically indistinguishable \cite{Bruno_2008} for all alloys with the same composition.
Hence the same applies for the quantities $b_X$, $a_X$ and $E_0$: these quantities are transferable between such systems.

The complexity of the CEFM is significantly reduced if one makes the assumption that $a_X$ takes the same value $a$ for all 
species \cite{Drchal_2006,Bruno_2007,Underwood_2013}. This assumption is known as the non-random approximation (NRA), and we use it throughout this 
work. The non-random approximation is borne out calculations utilizing the single-site locally self-consistent Green's function method 
\cite{Ruban_2002} - a GCPA method. In the 
NRA-CEFM $Q_i$ for an $X$ site obeys \cite{Underwood_2013}
\begin{equation}\label{Qi}
Q_i=\Lambda\sum_Yb_{YX}\sum_{\beta=1}^{\infty}g_{\beta}N_{iY\beta},
\end{equation}
where: $N_{iY\beta}$ is the number of $Y$ sites in the $\beta$th nearest neighbor shell of site $i$; 
\begin{equation}\label{bYX_def}
b_{YX}\equiv(b_Y-b_X);
\end{equation}
and the quantities $\Lambda$ and $g_{\beta}$ for all $\beta$ depend only on $aR_{\text{WS}}$ and the underlying lattice \emph{type}
\footnote{By lattice type we mean, for example, fcc, bcc, sc. Note that two lattices with the same type can have different values
of $R_{\text{WS}}$.}
- where $R_{\text{WS}}$ is the Wigner-Seitz radius for the system under consideration - and are tabulated in 
Ref. \onlinecite{Underwood_2013} for the fcc, bcc and sc lattices. Note that the free parameters $a$ and $b_{YX}$ (for all $X,Y$)
can be obtained from \emph{ab initio} calculations or by other means \cite{Underwood_2013}. Furthermore, the values of $N_{iY\beta}$ 
for all $\beta$ and $Y$ characterize the environment of $i$. Therefore Eqn. \eqref{Qi} explicitly relates $Q_i$ to the environment of site $i$.
Eqn. \eqref{Qi} also allows us to interpret an alloy's charge distribution in
terms of charge transfer between pairs of \emph{unlike} sites as follows \cite{Underwood_2013}: an $X$ site gains a charge 
$\Lambda b_{YX}g_{\beta}$ from each $Y$ site in its $\beta$th nearest neighbor shell, with the $Y$ site losing the opposite amount. This 
picture allows us to attribute the following physical significance to the quantity $b_Y$: it is a measure of the \emph{electropositivity} of
species $Y$ for the given composition.

\subsection{Expression for $\Delta E^{\text{B}}_i$}\label{sec:NRACEFM_DeltaEBi}
We will now use the NRA-CEFM to derive an expression for the CLS of site $i$. To do this, we first derive an expression for the total energy $E$ in terms of the bare charges $b_i$ of 
all sites. Eqn. \eqref{E_gen_def} can be rewritten more explicitly as
\begin{equation}\label{E_gen_gen_def_2}
E=E_0+\frac{1}{2}a\sum_i(Q_i-b_i)^2+\frac{1}{2}\sum_iQ_iV_i
\end{equation}
for the NRA-CEFM, where we have used Eqn. \eqref{EM_def}. Minimizing this with respect to the site charges, and subject to the constraint of global charge neutrality
gives \cite{Underwood_2013}
\begin{equation}
V_i=-aQ_i+a\Bigl(b_i-\langle b\rangle\Bigr),
\end{equation}
where $\langle b\rangle$ denotes the mean value of $b_i$ over all $i$.
Substituting the above into Eqn. \eqref{E_gen_gen_def_2} and simplifying gives
\begin{equation}\label{proof_E}
E=\frac{1}{2}a\sum_i\bigl(b_i^2-b_iQ_i\bigr)+E_0,
\end{equation}
where we have used the fact that
\begin{equation}
\sum_iQ_i=0.
\end{equation}
Now, the following expression holds for the charges at the minimum in $E$ \cite{Underwood_2013}:
\begin{equation}\label{proof_Q}
Q_i=a\sum_jG_{ij}\Bigl(b_j-\langle b\rangle\Bigr),
\end{equation}
where
\begin{equation}
G=(aI+M)^{-1},
\end{equation}
$M$ denotes the Madelung matrix, and $I$ denotes the identity matrix. This becomes
\begin{equation}\label{proof_Q_2}
Q_i=a\sum_jG_{ij}b_j
\end{equation}
since \cite{Drchal_2006}
\begin{equation}
\sum_jG_{ij}=0.
\end{equation}
Substituting Eqn. \eqref{proof_Q_2} into Eqn. \eqref{proof_E} gives
\begin{equation}
E=\frac{1}{2}a\sum_ib_i^2-\frac{1}{2}a^2\sum_i\sum_jG_{ij}b_ib_j+E_0,
\end{equation}
which becomes
\begin{equation}\label{proof_E_2}
\begin{split}
E=&\frac{1}{2}ab_k^2 -\frac{1}{2}a^2G_{kk}b_k^2 -a^2b_k\sum_{j\neq k}G_{jk}b_j+E_0 \\
&+ \frac{1}{2}a\sum_{i\neq k}b_i^2-\frac{1}{2}a^2\sum_{i\neq k}\sum_{j\neq k}G_{ij}b_ib_j.
\end{split}
\end{equation}
after separating out the terms containing $b_k$ and noting that $G$ is a symmetric matrix \cite{Drchal_2006}.
We will use the above equation in a moment.

The binding energy of a core level associated with site $k$ is
\begin{equation}
E^{\text{B}}_k=E^{\text{f}}_k-E^{\text{i}},
\end{equation}
where $E^{\text{i}}$ denotes the energy of the alloy's initial state - before photoemission from site $k$, and
$E^{\text{f}}_k$ denotes the energy of the alloy's final state - after photoemission from site $k$. In the 
complete screening picture \cite{Johansson_1980} the valence electrons in the final state are assumed to be fully 
relaxed so to reach their minimum energy configuration. With this in mind, the energy of the final state is simply the energy
of the initial state, but with the atomic core within $k$ replaced by its photo-ionized analogue. We choose site $k$
to belong to species $X$, and denote the `species' corresponding to a photo-ionized $X$ site as $X^*$. Hence
$E^{\text{B}}_k$ is the energy change if site $k$, originally belonging to species $X$, is transformed into species $X^*$.
Eqn. \eqref{proof_E_2} allows us to evaluate the change in $E$ due to such a transformation. Noting that the terms on 
the lower line of Eqn. \eqref{proof_E_2} are unaffected by the transformation (since $a$ and $b_X$ for all $X$ are
composition-dependent, and the composition - which we defined in terms of macroscopic quantities - is unaffected by the transformation),
that the transformation is such that $b_k^2\to b_{X^*}^2=b_X^2+2b_Xb_{X^*X} +b_{X^*X}^2$, and also that 
$E_0\to E_0+(\mathcal{E}_{X^*}-\mathcal{E}_X)$ (see Eqn. \eqref{E0_def}), we obtain
\begin{equation}
\begin{split}
E^{\text{B}}_k=&ab_Xb_{X^*X}+ \frac{1}{2}ab_{X^*X}^2- a^2G_{kk}b_Xb_{X^*X} \\
&- \frac{1}{2}a^2G_{kk}b_{X^*X}^2 - a^2b_{X^*X}\sum_{j\neq k}G_{jk}b_j \\
&+ (\mathcal{E}_{X^*}-\mathcal{E}_X).
\end{split}
\end{equation}
This can be rearranged to give
\begin{equation}
\begin{split}
E^{\text{B}}_k=&\frac{1}{2}a(1-aG_{kk})b_{X^*X}^2 +ab_Xb_{X^*X} \\
&- a^2b_{X^*X}\sum_jG_{jk}b_j  + (\mathcal{E}_{X^*}-\mathcal{E}_X),
\end{split}
\end{equation}
which in turn becomes
\begin{equation}\label{EBk}
\begin{split}
E^{\text{B}}_k=&- ab_{X^*X}Q_k  \\
& + \frac{1}{2}a(1-\Lambda)b_{X^*X}^2 +ab_Xb_{X^*X}  + (\mathcal{E}_{X^*}-\mathcal{E}_X)
\end{split}
\end{equation}
after using Eqn. \eqref{proof_Q_2} and noting that $\Lambda\equiv aG_{kk}$ \cite{Underwood_2013}. We emphasize that $Q_k$ in the above 
equation refers to the charge of site $k$ \emph{before} photoemission. 
Finally, substituting the above equation into Eqn. \eqref{CLS_def}, and relabeling site $k$ as site $i$, we obtain the following expression
for the CLS associated with site $i$:
\begin{equation}\label{DeltaEBk}
\Delta E^{\text{B}}_i=-ab_{X^*X}Q_i+\Phi_X,
\end{equation}
where
\begin{equation}
\begin{split}
\Phi_X\equiv& \frac{1}{2}a(1-\Lambda)b_{X^*X}^2 +ab_Xb_{X^*X} \\
& +(\mathcal{E}_{X^*}-\mathcal{E}_X) -E^{\text{B}}_{X\text{met}}
\end{split}
\end{equation}
is composition-dependent. 

Eqn. \eqref{DeltaEBk} describes a species- and composition-dependent linear mapping between $Q_i$ and $\Delta E^{\text{B}}_i$. 
Therefore changes in the $X$ CLS distribution associated with configuration changes - such as an increase in the interface roughening $\sigma$ in a 
multilayer system - directly reflect changes in the $X$ charge distribution. Furthermore the \emph{shape} of the CLS distribution
is the same as that of the charge distribution on account of the linear nature of the mapping between $Q_i$ and $\Delta E^{\text{B}}_i$. We
will elaborate on these points later.

\subsection{Mean and FWHM of various CLS distributions}\label{sec:mean_FWHM}
Henceforth we consider the mean and FWHM of the CLS distributions for various groups of $X$ sites.
Note that the FWHM of any random variable $x$ is related to its variance by the equation 
\begin{equation}\label{FWHM_def}
\FWHM(x)=2\sqrt{2\ln(2)}\sqrt{\Var(x)}.
\end{equation}
We will use this fact several times below. We will also use the fact that the mean and FWHM of $\Delta E^{\text{B}}_i$ for \emph{any group} of $X$ sites 
$S$ within a given alloy, as follows from Eqn. \eqref{DeltaEBk}, are given by
\begin{equation}\label{DeltaEBX_gen}
\langle\Delta E^{\text{B}}\rangle_X^S=-ab_{X^*X}\langle Q\rangle_X^S+\Phi_X
\end{equation}
and
\begin{equation}\label{FWHM_DeltaEBX_gen}
\Gamma_X^S=2\sqrt{2\ln(2)}\;|ab_{X^*X}|\sqrt{\Var(Q)_X^S}
\end{equation}
respectively, where $\langle Q\rangle_X^S$ and $\Var(Q)_X^S$ denote the mean and variance of the charge distribution for $S$.

\subsubsection{Random alloys}
Consider the group of all sites within a random alloy. For random alloys \cite{Underwood_2013}
\begin{equation}
\langle Q\rangle_X=-\Lambda\sum_Yb_{YX}c_Y
\end{equation}
and
\begin{equation}
\Var(Q)_X=\Lambda^2\omega\Var(b),
\end{equation}
where $\Var(b)$ denotes the variance of $b_i$ over \emph{all sites in the system}, and $\omega$ depends only on $aR_{\text{WS}}$ and
the underlying lattice type, and is tabulated in Ref. \onlinecite{Underwood_2013} for the fcc, bcc and sc lattices.
Therefore for random alloys Eqns. \eqref{DeltaEBX_gen} and \eqref{FWHM_DeltaEBX_gen} yield
\begin{equation}
\langle\Delta E^{\text{B}}\rangle_X=a\Lambda \sum_Y\tilde{b}_{YX}c_Y+\Phi_X
\end{equation}
and
\begin{equation}
\begin{split}
\Gamma_X=&2\sqrt{2\ln(2)}\;|a\Lambda\sqrt{\omega}| \\
&\times\sqrt{\sum_Y\tilde{b}_{YX}^2c_Y-\biggl(\sum_Y\tilde{b}_{YX}c_Y\biggr)^2},
\end{split}
\end{equation}
where we have defined
\begin{equation}
\tilde{b}_{YX}\equiv b_{X^*X}b_{YX}
\end{equation}
and used the fact that \cite{Underwood_2013}
\begin{equation}
\Var(b)=\sum_Yb_{YX}^2c_Y-\biggl(\sum_Yb_{YX}c_Y\biggr)^2.
\end{equation}

For binary alloys consisting of species $A$ and $B$ the above equations simplify to
\begin{equation}
\langle\Delta E^{\text{B}}\rangle_A=a\Lambda \tilde{b}_{BA}c_B+\Phi_A,
\end{equation}
and
\begin{equation}\label{FWHM_bin_rand}
\Gamma_A=2\sqrt{2\ln(2)}\;|a\Lambda\sqrt{\omega}\;\tilde{b}_{BA}|\sqrt{c_B(1-c_B)}
\end{equation}
for $X=A$.

\subsubsection{Mean for a single monolayer}
Henceforth we consider the class of systems described at the beginning of this section in which the species concentrations

can vary from monolayer to monolayer.
Let $S$ be the set of $X$ sites within monolayer $l$. Furthermore, let $\langle \Delta E^{\text{B}}\rangle_X^l$ and $\Gamma_X^l$
denote the mean and FWHM respectively of the CLS distribution for $S$. A similar notation will be used later for other quantities, e.g. 
$\langle Q\rangle_X^l$ and $\Var(Q)_X^l$.
We will now derive an expression for $\langle \Delta E^{\text{B}}\rangle_X^l$. Consider a site $i$ within $S$. From Eqn. \eqref{Qi}, $Q_i$ 
can be expressed as
\begin{equation}\label{Qi_layer_resolved}
Q_i=\Lambda\sum_Yb_{YX}\sum_{\beta=1}^{\infty}g_{\beta}\sum_mN_{iY\beta}^m,
\end{equation}
where $N_{iY\beta}^m$ denotes the number of $Y$ sites in the $\beta$th nearest neighbor shell of $i$ which are in monolayer $m$.
Taking the mean over all $i\in S$ gives
\begin{equation}
\langle Q\rangle_X^l=\Lambda\sum_Yb_{YX}\sum_{\beta=1}^{\infty}g_{\beta}\sum_m\langle N_{Y\beta}^m\rangle_X^l.
\end{equation}
Now, $N_{iY\beta}^m$ over $i\in S$ describes a random variable distributed according to the multinomial distribution. Specifically,
$N_{iY\beta}^m$ is the number of times outcome $Y$ occurs in $Z_{\beta}^{|l-m|}$ trials, given the probability of outcome $Y$ in a single
trial is $c_Y^m$, where $Z_{\beta}^d$ is the total number of sites in the $\beta$th nearest neighbor shell of any site $j$ which also 
belong to \emph{one} monolayer which is `$d$ monolayers away' from $j$ - with $d=0$ referring to the monolayer which contains site $j$ 
itself. The properties of the multinomial distribution are such that $\langle N_{Y\beta}^m\rangle_X^l=Z_{\beta}^{|l-m|}c_Y^m$, and
hence the above equation becomes
\begin{equation}
\langle Q\rangle_X^l=\Lambda\sum_Yb_{YX}\sum_{\beta=1}^{\infty}g_{\beta}\sum_mZ_{\beta}^{|l-m|}c_Y^m.
\end{equation}
This can be rewritten as
\begin{equation}\label{QXl}
\langle Q\rangle_X^l=\frac{1}{aR_{\text{WS}}}\sum_Yb_{YX}\sum_m\alpha^{|l-m|}c_Y^m,
\end{equation}
where we have defined
\begin{equation}\label{alpha_Def}
\alpha^d\equiv aR_{\text{WS}}\Lambda\sum_{\beta=1}^{\infty}g_{\beta}Z_{\beta}^d.
\end{equation}
Finally, substituting Eqn. \eqref{QXl} into Eqn. \eqref{DeltaEBX_gen} gives
\begin{equation}\label{mean_layer_CLS}
\langle\Delta E^{\text{B}}\rangle^l_X=-\frac{1}{R_{\text{WS}}}\sum_Y\tilde{b}_{YX}\sum_m\alpha^{|l-m|}c_Y^m+\Phi_X.
\end{equation}

For species $A$ in a binary alloy the above simplifies to
\begin{equation}
\langle\Delta E^{\text{B}}\rangle^l_A=-\frac{\tilde{b}_{BA}}{R_{\text{WS}}}\sum_m\alpha^{|l-m|}c_B^m+\Phi_B.
\end{equation}

The quantities $\alpha^d$ depend on the lattice type, $aR_{\text{WS}}$, and the set of lattice planes which constitute the monolayers. They
determine the coupling between monolayers with regards to $\langle Q\rangle_X^l$ and $\langle \Delta E^{\text{B}}\rangle_X^l$.
It can be shown that for the concentration profile $c_X^l=c_X$ for all $l$, i.e., a random alloy, the above equations become
equivalent to those given earlier for random alloys. This is the case since
\begin{equation}
\sum_m\alpha^{|l-m|}=-aR_{\text{WS}}\Lambda,
\end{equation}
which can be shown by noting that
\begin{equation}\label{Zbeta_sum}
Z_{\beta}=\sum_mZ_{\beta}^m
\end{equation}
and \cite{Underwood_2013}
\begin{equation}
\sum_{\beta=1}^{\infty}g_{\beta}Z_{\beta}=-1.
\end{equation}

\subsubsection{FWHM for a single monolayer}
We will now derive an expression for $\Gamma_X^l$. Taking the variance of Eqn. \eqref{Qi_layer_resolved} over
$i\in S$ gives
\begin{equation}
\Var(Q)^l_X=\Lambda^2\sum_{\beta=1}^{\infty}g_{\beta}^2\sum_m\Var\Biggl(\sum_Yb_{YX}N_{Y\beta}^m\Biggr)^l_X
\end{equation}
after noting that the random variables $N_{Y\beta}^m$ and $N_{Z\gamma}^n$, defined by considering $N_{iY\beta}^m$ and
$N_{iZ\gamma}^n$ for $i\in S$, are independent if $\gamma\neq\beta$ or $n\neq m$. Expanding the variance on the 
right-hand side gives
\begin{equation}\label{VarQlX_layer}
\begin{split}
\Var(Q)^l_X=&\Lambda^2\sum_{\beta=1}^{\infty}g_{\beta}^2\sum_m\Biggl[
\sum_Yb_{YX}^2\Var(N_{Y\beta}^m)^l_X \\
&+\sum_Y\sum_{Z\neq Y}b_{YX}b_{ZX}\Cov(N_{Y\beta}^m,N_{Z\beta}^m)^l_X\Biggr],
\end{split}
\end{equation}
where $\Cov(x,y)$ denotes the covariance of random variables $x$ and $y$. Because $N_{iY\beta}^m$ over $i\in S$ form
a multinomial distribution,
\begin{equation}
\Var(N_{Y\beta}^m)^l_X=Z_{\beta}^{|l-m|}c_Y^m(1-c_Y^m)
\end{equation}
and
\begin{equation}
\Cov(N_{Y\beta}^m,N_{Z\beta}^m)^l_X=-Z_{\beta}^{|l-m|}c_Y^mc_Z^m\quad\text{if $Z\neq Y$}.
\end{equation}
Substituting these equations into Eqn. \eqref{VarQlX_layer} gives
\begin{equation}\label{VarQlX_layer_2}
\begin{split}
\Var(Q)^l_X=&\frac{1}{(aR_{\text{WS}})^2}\sum_m\beta^{|m-l|}\Biggl[\sum_Yb_{YX}^2c_Y^m(1-c_Y^m) \\
&-\sum_Y\sum_{Z\neq Y}b_{YX}b_{ZX}c_Y^mc_Z^m\Biggr],
\end{split}
\end{equation}
where we have defined
\begin{equation}\label{beta_def}
\beta^d\equiv (aR_{\text{WS}})^2\Lambda^2\sum_{\beta=1}^{\infty}g_{\beta}^2Z_{\beta}^d.
\end{equation}
Substituting Eqn. \eqref{VarQlX_layer_2} into Eqn. \eqref{FWHM_DeltaEBX_gen} gives
\begin{widetext}
\begin{equation}\label{FWHM_layer_CLS}
\Gamma^l_X=2\sqrt{2\ln(2)}\;\frac{1}{R_{\text{WS}}}
\sqrt{\sum_m\beta^{|m-l|}\Biggl[\sum_Y\tilde{b}_{YX}^2c_Y^m(1-c_Y^m)-\sum_Y\sum_{Z\neq Y}\tilde{b}_{YX}\tilde{b}_{ZX}c_Y^mc_Z^m\Biggr]}
\end{equation}
\end{widetext}

For species $A$ in a binary alloy the above equation simplifies significantly:
\begin{equation}
\Gamma^l_A=2\sqrt{2\ln(2)}\;\frac{|\tilde{b}_{BA}|}{R_{\text{WS}}}\sqrt{\sum_m\beta^{|m-l|}c_B^m(1-c_B^m)}.
\end{equation}

The quantities $\beta^d$ determine the coupling between monolayers with regards to $\Var(Q)^l_X$ and $\Gamma^l_X$.
As was the case for $\langle\Delta E^{\text{B}}\rangle^l_X$, if $c_X^l=c_X$
for all $l$ then the above equations become equivalent to those given earlier for random alloys. This is the case since
\begin{equation}
\sum_m\beta^{|l-m|}=(aR_{\text{WS}})^2\Lambda^2\omega,
\end{equation}
which can be shown by appealing to Eqn. \eqref{Zbeta_sum} as well as the definition of
$\omega$:
\begin{equation}
\omega\equiv\sum_{\beta=1}^{\infty}g_{\beta}^2Z_{\beta}.
\end{equation}

\subsubsection{Total mean}
From the quantities $\langle\Delta E^{\text{B}}\rangle^l_X$ for all $l$, the mean CLS $\langle\Delta E^{\text{B}}\rangle_X$ over \emph{all} $X$ sites in
the system under consideration can be determined by using the following expression:
\begin{equation}\label{tot_mean_CLS}
\langle\Delta E^{\text{B}}\rangle_X=\sum_lw_X^l\langle\Delta E^{\text{B}}\rangle_X^l,
\end{equation}
where $w_X^l$ is the \emph{weight} to be given to monolayer $l$. For the `true' mean, $w_X^l$ is given by
\begin{equation}
w_X^l=c_X^l\bigg/\sum_mc_X^m.
\end{equation}
However, one is often interested in the mean of the CLS distribution observed experimentally, which may differ from the `true' value on 
account of the fact that the probability $p^l$ of a photoelectron emitted from monolayer $l$ escaping the alloy is monolayer-dependent.
In this case $w_X^l$ is given by
\begin{equation}
w_X^l=p^lc_X^l\bigg/\sum_mp^mc_X^m.
\end{equation}

Eqn. \eqref{tot_mean_CLS} follows trivially from the following theorem: if $\Sigma_i$ is a set of $n_i$ values whose mean is $\mu_i$, then 
the mean of the superset $\Sigma$ formed by combining the sets $\Sigma_i$ for all $i$ is
\begin{equation}\label{mean_theorem}
\mu=\sum_i\frac{n_i}{n}\mu_i.
\end{equation}
This result can be derived from the definitions of $\mu$ and $\mu_i$:
\begin{equation}
\mu=\frac{1}{n}\sum_{\varepsilon\in\Sigma}\varepsilon=\frac{1}{n}\sum_i\sum_{\varepsilon\in\Sigma_i}\varepsilon
=\sum_i\frac{n_i}{n}\Biggl(\frac{1}{n_i}\sum_{\varepsilon\in\Sigma_i}\varepsilon\Biggr).
\end{equation}

\subsubsection{Total FWHM}
From the quantities $\langle\Delta E^{\text{B}}\rangle_X$, and $\langle\Delta E^{\text{B}}\rangle^l_X$ and $\Gamma^l_X$ for all $l$,
the FWHM in the CLS distribution over all $X$ sites $\Gamma_X$ can determined using the following expression:
\begin{widetext}
\begin{equation}\label{tot_FWHM_CLS}
\Gamma_X=\sqrt{\sum_lw_X^l\biggl[\Bigl(\Gamma_X^l\Bigr)^2
+8\ln2\Bigl(\langle\Delta E^{\text{B}}\rangle_X^l-\langle\Delta E^{\text{B}}\rangle_X\Bigr)^2\biggr]}.
\end{equation}
\end{widetext}

Eqn. \eqref{tot_FWHM_CLS} can be derived by applying Eqn. \eqref{FWHM_def} to the analogous equation to Eqn. \eqref{mean_theorem} for the 
variance:
\begin{equation}\label{FWHM_theorem}
\nu=\sum_i\frac{n_i}{n}\Bigl[\nu_i+(\mu_i-\mu)^2\Bigr],
\end{equation}
where $\nu_i$ denotes the variance of $\Sigma_i$, and $\nu$ denotes the variance of $\Sigma$. The above equation can itself be derived somewhat
similarly to Eqn. \eqref{mean_theorem}:
\begin{equation}
\begin{split}
\nu=&\frac{1}{n}\sum_{\varepsilon\in\Sigma}(\varepsilon-\mu)^2=\sum_i\frac{n_i}{n}\frac{1}{n_i}\sum_{\varepsilon\in\Sigma_i}(\varepsilon-\mu)^2 \\
=&\sum_i\frac{n_i}{n}\frac{1}{n_i}\sum_{\varepsilon\in\Sigma_i}\Bigl[(\varepsilon-\mu_i)+(\mu_i-\mu)\Bigr]^2 \\
=&\sum_i\frac{n_i}{n}\Biggl[\frac{1}{n_i}\sum_{\varepsilon\in\Sigma_i}(\varepsilon-\mu_i)^2+(\mu_i-\mu)^2 \\
&+2\frac{1}{n_i}(\mu_i-\mu)\sum_{\varepsilon\in\Sigma_i}(\varepsilon-\mu_i)\Biggr],
\end{split}
\end{equation}
which becomes Eqn. \eqref{FWHM_theorem} after noting that the final term in the last equality vanishes.

\section{Computational details}\label{sec:comp_details}
\begin{figure}
\centering
\includegraphics[width=0.5\textwidth]{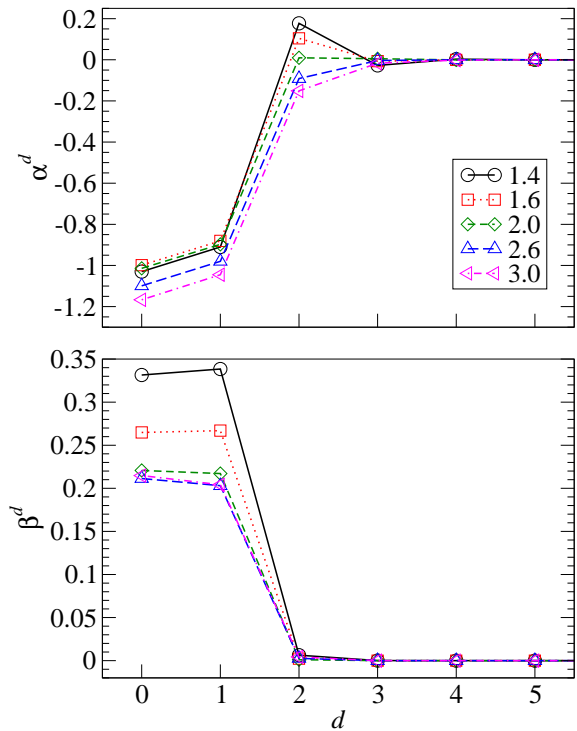}
\caption{(Color online) The monolayer coupling parameters $\alpha^d$ and $\beta^d$ for the 001 planes in the fcc lattice for various values of 
$aR_{\text{WS}}$. The top panel shows $\alpha^d$ vs. $d$; while the bottom panel shows $\beta^d$ vs. $d$. The $aR_{\text{WS}}$
to which each curve corresponds can be deduced from the key in the top panel.}
\label{fig:structural_params}
\end{figure}

In the next section we apply the expressions of Sec. \ref{sec:mean_FWHM} to various multilayer systems. All systems we consider have
an fcc underlying lattice, and the monolayers are the 001 planes.
However, to apply these expressions we first had to determine the monolayer coupling parameters $\alpha^d$ 
and $\beta^d$, which are defined in Eqns. \eqref{alpha_Def} and \eqref{beta_def}. In Ref. \onlinecite{Underwood_2013}, $\Lambda$ and $g_{\beta}$ 
are tabulated as a 
function of $aR_{\text{WS}}$ for the fcc, bcc and sc lattices. Using this information, and after determining the quantities $Z_{\beta}^d$ - which 
depend on the underlying geometry - one can tabulate $\alpha^d$ and $\beta^d$ as a function of $aR_{\text{WS}}$ for any fcc, bcc or sc system. 
For other lattice types, $\Lambda$ and $g_{\beta}$ must be determined as a function of $aR_{\text{WS}}$ beforehand. A procedure to do 
this is described in Ref. \onlinecite{Underwood_2013}. 
We determined $\alpha^d$ and $\beta^d$ for the 001 planes of the fcc lattice at selected $aR_{\text{WS}}$.
The results are shown in Fig. \ref{fig:structural_params}. We considered $aR_{\text{WS}}=$1.4, 1.6, 2.0, 2.6 and 3.0, which reflects the
range of $aR_{\text{WS}}$ found in the literature \cite{Faulkner_1995,Faulkner_1997,Ruban_2002,Bruno_2008,Bruno_2002,Ruban_2002_2,Olovsson_2011}.
Note that the coupling between monolayers rapidly tends to zero with $d$.

In addition to the aforementioned `analytical' results, we also performed supercell calculations for each of the systems we considered. In 
our supercells, each monolayer contained 200 sites, with the species of sites in each monolayer $l$ assigned 
randomly such that the desired species concentrations $c_X^l$ were obtained as closely as possible. 
$\Delta E^{\text{B}}_i$ was determined for each site using Eqns. 
\eqref{DeltaEBk} and \eqref{Qi}. These $\Delta E^{\text{B}}_i$ where then used to determine $\langle\Delta E^{\text{B}}\rangle_X^l$, 
$\Gamma^l_X$, $\langle\Delta E^{\text{B}}\rangle_X$ and $\Gamma_X$ in the conventional manner, which allowed us to cross-check our analytical results. 
For all systems the analytical and supercell results were in excellent agreement, though we choose not to present the supercell means and FWHMs
for the sake of brevity and clarity of presentation. In addition to cross-checking, the supercell values of $\Delta E^{\text{B}}_i$ were used to 
simulate core level spectra for some systems. In this regard we used the following equation:
\begin{equation}\label{spectrum_exact}
I(\Delta E^{\text{B}})=\sum_{i\in X}L\Bigl(\Delta E^{\text{B}}-\Delta E^{\text{B}}_i;\Gamma_{\text{life}}\Bigr),
\end{equation}
where $I(\Delta E^{\text{B}})$ denotes the intensity of the $X$ spectrum at CLS $\Delta E^{\text{B}}$, $L(\Delta E^{\text{B}};\Gamma)$ is a 
Lorentzian function with FWHM $\Gamma$, and $\Gamma_{\text{life}}$ is the lifetime broadening of the core levels
under consideration. The above equation does not take into account many features which are present in `real' spectra such as 
surface core-level shifts, experimental broadening, Doniac-Sunjic asymmetry, and inelastic scattering. However, since our simulated spectra
are primarily for illustrative purposes, ignoring these complications is justified. Our supercell calculations served one further purpose.
Later we present histograms of, for binary systems, the frequency of $A$ sites exhibiting each possible number of $B$ nearest neighbors, and
for ternary systems, the frequency of $A$ sites exhibiting each possible combination of $B$ and $C$ nearest neighbors. We used our supercells
to generate these histograms: for each supercell we counted the number of $A$ sites with each possible composition of nearest 
neighbor shell.

\section{Results}\label{sec:results}

\subsection{Isolated embedded thin film: $B/A_T/B$}
We first present results for systems consisting of a thin film of species $A$ and thickness $T$ monolayers embedded in a $B$ substrate, i.e., $B/A_T/B$. We 
examined such systems with various $T$ and $\sigma$, where recall that $\sigma$ denotes the degree of interface roughening. 
Following Refs. \onlinecite{Holmstrom_2004,Olovsson_2011}, interface roughening was modeled by convoluting the `unroughened' concentration profile 
with a discrete Gaussian function \cite{Holmstrom_2004}, where $\sigma$ denotes the standard deviation of the function.
To elaborate, the concentration profile corresponding to $\sigma$ was calculated using the following equation:
\begin{equation}\label{conc_sigma}
c^l_X(\sigma)=\sum_m\gamma^{l-m}(\sigma)c^l_X(0)
\end{equation}
for all $X$, where 
\begin{equation}
\gamma^{d}(\sigma)=\frac{\exp\big[-(d/\sigma)^2/2\bigr]}{\displaystyle\sum_{d'=-\infty}^{\infty}\exp\big[-(d'/\sigma)^2/2\bigr]}
\end{equation}
is the discrete analogue of a Gaussian function with standard deviation $\sigma$, and $c^l_X(0)$ denotes the unroughened concentration profile.

\subsubsection{General results}
\begin{figure}
\centering
\includegraphics[width=0.5\textwidth]{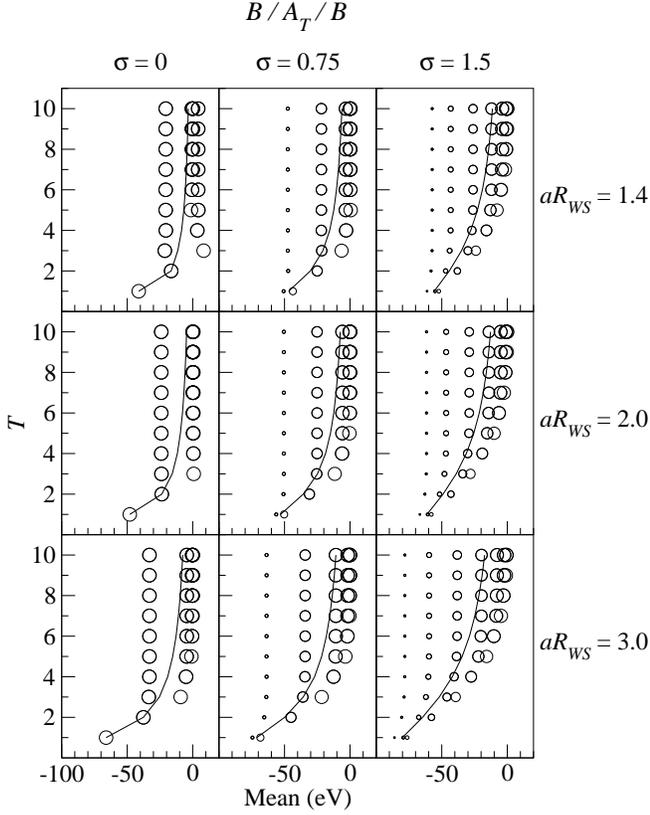}
\caption
{Mean CLSs for $B/A_T/B$ systems calculated using the model. Each column pertains to a different value of $\sigma$, which is indicated above the
column; each row pertains to a different value of $aR_{\text{WS}}$, which is indicated to the right of the row. Each circle represents 
$\langle\Delta E^{\text{B}}\rangle_A^l$ for a particular monolayer $l$; the size of the symbol is proportional to $c_A^l$. Only results pertaining 
to monolayers with $c_A^l>0.05$ are shown. The solid lines connect $\langle\Delta E^{\text{B}}\rangle_A$ for each value of $T$.}
\label{fig:mean_BAB}
\end{figure}

\begin{figure}
\centering
\includegraphics[width=0.5\textwidth]{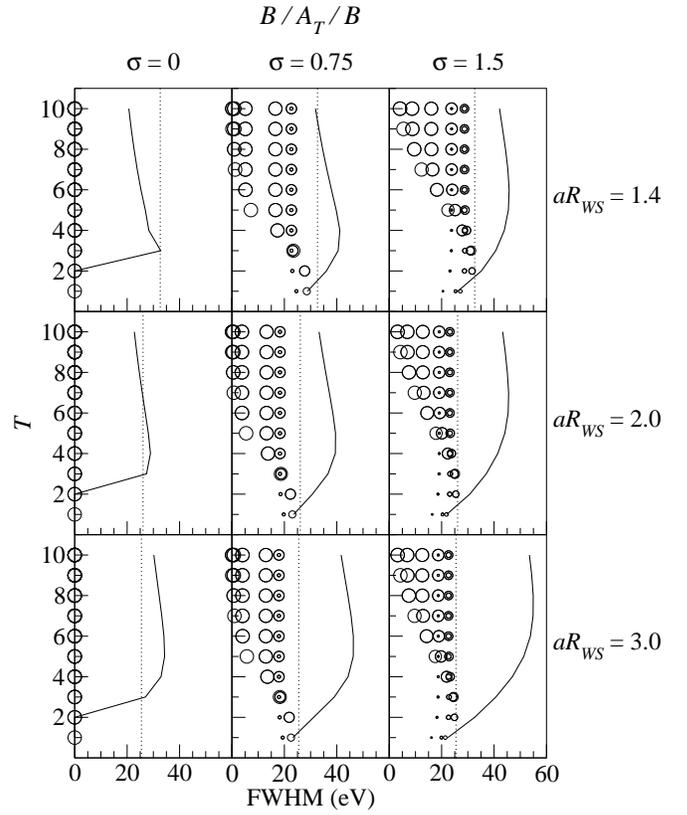}
\caption
{FWHM CLSs for $B/A_T/B$ systems calculated using the model. The details of the figure are the same as in the caption to Fig. 
\eqref{fig:mean_BAB}, except that circles represent $\Gamma^l_A$, and solid lines connect $\Gamma_A$ for each value of $T$. Furthermore, 
the dotted line in each panel corresponds to the upper bound on the FWHM of a random alloy at the corresponding value of $aR_{\text{WS}}$.}
\label{fig:fwhm_BAB}
\end{figure}

The quantities $\langle\Delta E^{\text{B}}\rangle_A^l$, $\langle\Delta E^{\text{B}}\rangle_A$, $\Gamma^l_A$ and $\Gamma_A$ for various
$B/A_T/B$ systems are presented in Figs. \ref{fig:mean_BAB} and \ref{fig:fwhm_BAB}. In calculating these values, we used the set of
$aR_{\text{WS}}$ described in the previous section. Furthermore, we set $R_{\text{WS}}=1$, $\tilde{b}_{BA}\equiv b_{A^*A}b_{BA}=-1$, and $\Phi_A=0$. 
Results (in eV) obtained using these parameters can be generalized to any choice of $R_{\text{WS}}$ (in bohrs), $b_{A^*A}$, $b_{BA}$ (both in units of 
$e$) and $\Phi_A$ (in eV) at the corresponding value of $aR_{\text{WS}}$ by multiplying all means and FWHMs by $-b_{A^*A}b_{BA}/R_{\text{WS}}$, and additionally 
adding $\Phi_A$ to all mean CLSs. We will use this fact in a moment. 

Some interesting results are immediately apparent from Figs. \ref{fig:mean_BAB} and \ref{fig:fwhm_BAB}.
Firstly, as can be seen from both figures, the model results only depend weakly on $aR_{\text{WS}}$. This is convenient because the `true' value of 
$aR_{\text{WS}}$ for a given system is unclear; values of $aR_{\text{WS}}$ are very sensitive to the \emph{ab initio} method used to obtain them.
The same is true for the quantities $b_{YX}$, and presumably also for the quantities $b_{X^*X}$ and $\Phi_X$. This is illustrated in Table 
\ref{table:ab_initio_compare}, where the $aR_{\text{WS}}$ and $b_{\text{CuZn}}$ for the bcc random alloy Cu$_{0.5}$Zn$_{0.5}$ obtained using different 
\emph{ab initio} methods are compared. In calculating each $aR_{\text{WS}}$ in the table we took $a$ to be the mean of the \emph{ab initio}
\emph{Q-V} relation gradients $a_{\text{Cu}}$ and $a_{\text{Zn}}$ for Cu and Zn. Furthermore, we calculated $b_{\text{CuZn}}$ from the \emph{Q-V} relation
intercepts via $b_{\text{CuZn}}=(k_{\text{Cu}}-k_{\text{Zn}})/a$ (see Eqn. \eqref{k_def}).

\begin{table}
\begin{ruledtabular}
\begin{tabular}{ccc}
Method & $aR_{\text{WS}}$ & $b_{\text{CuZn}}$ \\
\hline
LSMS\cite{Faulkner_1995,Faulkner_1997,Bruno_2002} & 2.5 &  0.16 \\  
GCPA\cite{Bruno_2008} & 1.6 & 0.12 \\  
LAPW\cite{Bruno_2008} & 3.5 & 0.16 \\  
\end{tabular}
\end{ruledtabular}
\caption
{$aR_{\text{WS}}$ and $b_{\text{CuZn}}$ for Cu$_{0.5}$Zn$_{0.5}$ obtained using various \emph{ab initio} calculations. The abbreviation `LSMS' refers to the 
locally self-consistent Green's function method; and `LAPW' refers to the linearized augmented plane wave method. See Refs. 
\onlinecite{Faulkner_1995,Faulkner_1997,Bruno_2008} and references therein for details regarding these calculations.}
\label{table:ab_initio_compare}
\end{table}

Our second observation relates to Fig. \ref{fig:fwhm_BAB}. One might expect that, at a given composition and parametrization of the model (i.e.,
choice of the quantities $aR_{\text{WS}}$, $b_{A^*A}$, $b_{BA}$ and $\Phi_A$),
the $A$ FWHM would be maximized at the random alloy configuration. This configuration corresponds to the largest configurational entropy, and therefore might
be expected to exhibit the largest range of environments and hence also the largest FWHM. Moreover, since, as can be seen from Eqn. \eqref{FWHM_bin_rand}, 
$\sqrt{c_B(1-c_B)}$ is maximized when $c_A=c_B=0.5$, one might therefore also expect that the corresponding FWHM, which is an upper bound for the
FWHM of a random alloy, is also an upper bound for \emph{all} alloys. However, our results reveal that this is \emph{not} the case. 
In each panel of Fig. \ref{fig:fwhm_BAB}, the random alloy upper bound corresponding to the same model parametrization as we used for our 
$B/A_T/B$ systems is indicated by a dotted line.
Note that for some systems $\Gamma_A$ is significantly larger than the random alloy upper bound. It is even possible for the upper bound 
to be exceeded in `unroughened' ultra-thin films - as is evident from the left-most column in Fig. \ref{fig:fwhm_BAB}. Therefore
inhomogeneous concentration profiles can yield larger disorder broadenings than is possible in random alloys - a fact which we provide
an explanation for later. A similar observation has been made in our earlier study\cite{Underwood_2010} using the linear charge model\cite{Magri_1990}: 
surface segregation was shown to result in a significantly larger disorder broadening relative to the unsegregated random alloy. However in that study 
surface effects are implicit in the simulated spectra, and hence cannot be discounted as a contributing factor to the very large broadening. By contrast 
here we have shown that a very large broadening can occur in the absence of surface effects. In the aforementioned study we suggested that 
segregation could explain the anomalously large disorder broadening observed experimentally by Medicherla \emph{et al.}\cite{Medicherla_2009}. Our 
results here provide further evidence for this hypothesis. 

In contrast to $\Gamma_A$, the monolayer FWHMs $\Gamma_A^l$ seem to be constrained to be below the random alloy upper bound. 
This can be understood by noting that the $A$ sites in monolayer $l$ experience a local environment which closely resembles a random alloy, and hence
$\Gamma_A^l$ (to a good approximation) cannot exceed the upper bound. Considering
only nearest neighbors, each $A$ site in monolayer $l$ has $Z_1^0$ nearest neighbors in monolayer $l$, $Z_1^1$ in monolayer $l-1$, and $Z_1^1$ in 
monolayer $l+1$. Since \emph{within each monolayer} we have assigned sites' species randomly in the required concentrations, it follows that the
environment of an $A$ site in monolayer $l$ is approximately that of a random alloy with concentration
\begin{equation}
c_{A,\text{eff}}^l\approx Z_1^0c_A^l+Z_1^1c_A^{l-1}+Z_1^1c_A^{l+1}.
\end{equation}

\subsubsection{Comparison with \emph{ab initio} results}
\begin{figure}
\centering
\includegraphics[width=0.5\textwidth]{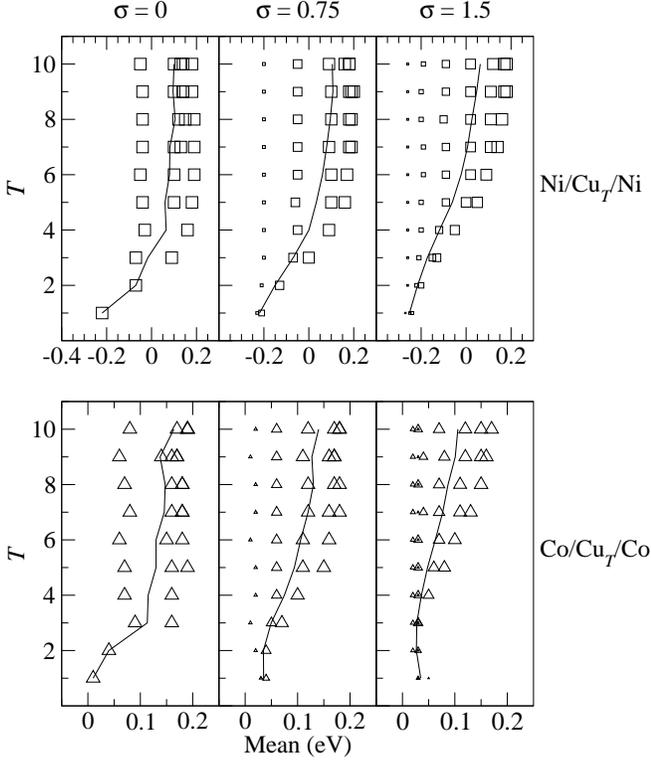}
\caption
{Mean CLSs for Ni/Cu$_T$/Ni and Co/Cu$_T$/Co systems calculated by Olovsson \emph{et al.} (Ref. \onlinecite{Olovsson_2011}).
The significance of each column, and the symbols and lines, is the same as in Fig. \ref{fig:mean_BAB}. The top row of graphs corresponds to 
Ni/Cu$_T$/Ni, while the bottom row corresponds to Co/Cu$_T$/Co.}
\label{fig:mean_Olovsson_BAB}
\end{figure}

The $B/A_T/B$ systems we considered were deliberately chosen to be identical to the fcc systems considered in Ref. \onlinecite{Olovsson_2011},
in which Olovsson \emph{et al.} calculated $\langle\Delta E^{\text{B}}\rangle_{\text{Cu}}^l$ for Ni/Cu$_T$/Ni and Co/Cu$_T$/Co. The 
$\langle\Delta E^{\text{B}}\rangle_{\text{Cu}}^l$ and $\langle\Delta E^{\text{B}}\rangle_{\text{Cu}}^l$ determined by Olovsson \emph{et al.} are 
shown in Fig. \ref{fig:mean_Olovsson_BAB}. Comparing this figure with Fig. \ref{fig:mean_BAB} we see that the model is in 
excellent qualitative agreement with those of Olovsson \emph{et al.} for the Ni/Cu$_T$/Ni systems. The agreement is especially good at 
higher values of $aR_{\text{WS}}$. For the Co/Cu$_T$/Co systems, the agreement is reasonable. However, it should be borne in mind that the
dispersion of CLSs in Co/Cu$_T$/Co is very small, and hence any uncertainties implicit in the \emph{ab initio} 
method used by Olovsson \emph{et al.} will be larger relative to the size of the dispersion: Olovsson \emph{et al.} quote CLSs 
to a precision of 10meV, which is significant on the scale of the Co/Cu$_T$/Co dispersion, but not for the Ni/Cu$_T$/Ni dispersion. 

The $R_{\text{WS}}$ used by Olovsson \emph{et al.} in their calculations was 2.6 bohrs for both Ni/Cu$_T$/Ni and Co/Cu$_T$/Co. With this in mind,
and recalling the procedure described above for generalizing our model results to different free parameters, we found that
$aR_{\text{WS}}\approx 2.6$, $\Phi_{\text{Cu}}\approx 0.15$eV and $b_{\text{Cu}^*\text{Cu}}b_{\text{NiCu}}\approx -0.015e^2$ gave excellent agreement with
Olovsson \emph{et al.} for Ni/Cu$_T$/Ni. For Co/Cu$_T$/Co we found acceptable agreement when $aR_{\text{WS}}\approx 2.6$, 
$b_{\text{Cu}^*\text{Cu}}b_{\text{CoCu}}\approx -0.01e^2$, and $\Phi_{\text{Cu}}\approx 0.2$eV.
These values are similar to analogous quantities obtained from \emph{ab initio} calculations for 
other alloys \cite{Faulkner_1995,Faulkner_1997,Ruban_2002,Bruno_2008,Bruno_2002,Ruban_2002_2,Olovsson_2011}.
However, as mentioned earlier, the values of these quantities are sensitive to the \emph{ab initio} method used to obtain them. Hence an 
interesting prospect is to use the model to determine their values experimentally.

Olovsson \emph{et al.} used the $\langle\Delta E^{\text{B}}\rangle_{\text{Cu}}^l$ from their calculations to simulate core level XPS spectra. However,
since their calculations did not provide values for $\Gamma_{\text{Cu}}^l$, it was necessary for Olovsson \emph{et al.} to make some assumptions regarding these 
quantities. They assumed that $\Gamma_{\text{Cu}}^l$ was the same for all $l$. Fig. \ref{fig:fwhm_BAB} reveals that, in fact, the quantities $\Gamma_{\text{Cu}}^l$
vary widely within any one system. Hence the assumption made by Olovsson \emph{et al.} is incorrect. However, it is 
unclear whether the breakdown of this assumption is important from a practical point of view. While accurate knowledge of the quantities $\Gamma_X^l$ 
- as well as perhaps higher moments of the $X$ CLS distribution for each monolayer - is necessary to reproduce the fine details of the total $X$ CLS
distribution for the system under consideration, in practice these fine details are `smeared out' in the experimental spectrum due to complications
such as lifetime and experimental broadening. Therefore getting some of the fine details wrong in the spectrum before accounting for the aforementioned
complications may be inconsequential with regards to accurately reproducing experimental spectra.

It should be noted that Olovsson \emph{et al.} considered a further system in their study in addition to Ni/Cu$_T$/Ni and Co/Cu$_T$/Co: Fe/Cu$_T$/Fe.
We do not perform a thorough comparison between our model calculations and those of Olovsson \emph{et al.} for Fe/Cu$_T$/Fe because the underlying
lattice for Fe/Cu$_T$/Fe is bcc, while our model calculations are for an fcc underlying lattice. 
At $\sigma=0.75$ and $\sigma=1.5$, the qualitative nature of the \emph{ab initio} Fe/Cu$_T$/Fe results is similar to that for Ni/Cu$_T$/Ni and Co/Cu$_T$/Co.
In these cases we therefore expect that the model will perform well for Fe/Cu$_T$/Fe. However, as pointed out by Olovsson \emph{et al.}, a 
well-known interface state exists in ordered, but not in disordered Fe/Cu$_T$/Fe systems, which causes `anomalous' results for $\sigma=0$. 
It would be interesting to see whether model calculations utilizing a bcc lattice can reproduce this.

\subsection{Periodic NiCu multilayers: [Ni$_T$/Cu$_U$]}
In the previous subsection we considered an isolated thin film embedded in an infinite substrate. This was done primarily to allow comparison with
analogous \emph{ab initio} results. However, experimental studies have focused on \emph{periodic} multilayer systems.
We now consider such systems; specifically, those comprised of Ni and Cu in which the repeating unit consists of $T$ monolayers of 
Ni adjacent to $U$ monolayers of Cu, i.e., [Ni$_T$/Cu$_U$]. We considered various $T$, $U$ and $\sigma$. As above, we used 
Eqn. \eqref{conc_sigma} to obtain the concentration profile for a given $\sigma$. With regards to the parametrization of 
the model, we used the `best-fit parametrization' described above for the Ni/Cu$_T$/Ni systems, i.e., $aR_{\text{WS}}=2.6$, $\Phi_{\text{Cu}}= 0.15$eV, and 
$b_{\text{Cu}^*\text{Cu}}b_{\text{NiCu}}=-0.015e^2$.

\begin{figure}
\centering
\includegraphics[height=0.5\textwidth,angle=270]{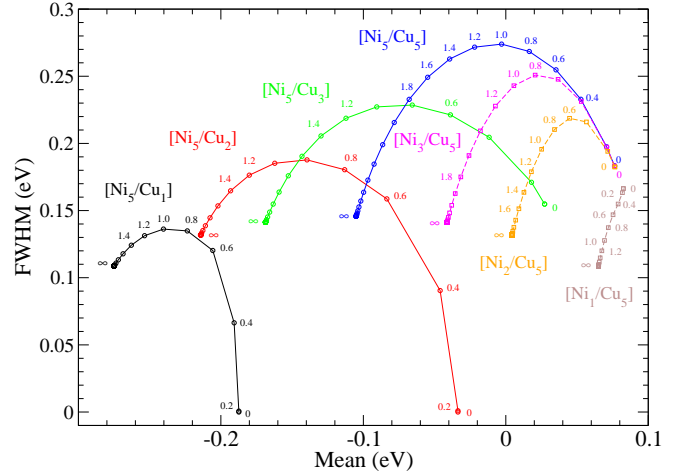}
\caption
{(Color online) Mean and FWHM of the Cu CLS distribution in various NiCu systems as a function of $\sigma$, determined using the model. Each color
pertains to a different system; each system is labeled. Symbols indicate the (mean,~FWHM) evaluated at $\sigma$ from 0 to $\infty$ at intervals of 
$\sigma$ of 0.2. Some of the symbols are annotated with their value of $\sigma$. Curves connect the symbols, and trace the (mean,~FWHM) from $\sigma=0$ to 
$\sigma=\infty$ for each system.}
\label{fig:trajectories_CuNi}
\end{figure}

The $\langle \Delta E^{\text{B}}\rangle_{\text{Cu}}$ and $\Gamma_{\text{Cu}}$ for various [Ni$_T$/Cu$_U$] systems as a function of $\sigma$ are presented 
in Fig. \ref{fig:trajectories_CuNi}; we show the `trajectory' of the pair $(\langle \Delta E^{\text{B}}\rangle_{\text{Cu}},\Gamma_{\text{Cu}})$
as $\sigma$ is varied from 0 to $\infty$. Note that at $\sigma=\infty$ intermixing between the Ni and Cu regions is absolute, and [Ni$_T$/Cu$_U$] becomes a 
random alloy with $c_{\text{Cu}}=U/(T+U)$.
As is evident from Fig. \ref{fig:trajectories_CuNi}, as $\sigma$ increases, $\langle \Delta E^{\text{B}}\rangle_{\text{Cu}}$ becomes more negative, and
$\Gamma_{\text{Cu}}$ increases initially, before decreasing and finally settling on the random alloy value. 
The exception is [Ni$_1$/Cu$_5$], for which $\Gamma_{\text{Cu}}$ monotonically decreases with $\sigma$, i.e., the system exhibits a `disorder narrowing'.
We will now explain these trends.

\subsubsection{Rationalization of spectral changes}
A pleasing feature of the model is that it provides a simple means of rationalizing changes in CLS spectra due to configurational
changes. Consider Eqn. \eqref{Qi}. An insightful approximation is to ignore the dependence of $Q_i$ on the environment of site $i$ beyond its nearest 
neighbor shell, in which case for Cu in NiCu systems the number of Ni nearest neighbors $N_{i\text{Ni}1}$ of site $i$ wholly determines $Q_i$ 
\footnote{A more accurate approach would be to apply `mean-field boundary conditions' beyond the nearest neighbor shell, i.e., assume that
$N_{i\text{Ni}\beta}=Z_{\beta}c_{\text{Ni}}$ for $\beta>1$, where recall that $Z_{\beta}$ denotes the number of sites in any site's $\beta$th nearest neighbor 
shell, and $c_{\text{Ni}}$ denotes the global concentration of Ni. This would lead to better agreement between the histograms and model spectra in Fig.
\ref{fig:figure_Cu5Ni5})(b). However, we do not do this here for the sake of simplicity.}.
Specifically, $Q_i$ is linear in $N_{i\text{Ni}1}$:
\begin{equation}\label{Q_nn_bin}
Q_i\approx\Lambda g_1b_{\text{NiCu}}N_{i\text{Ni}1}.
\end{equation}
The same then applies to $\Delta E^{\text{B}}_i$ (from Eqn. \eqref{DeltaEBk}):
\begin{equation}\label{DeltaEBi_nn_bin}
\Delta E^{\text{B}}_i\approx -a\Lambda g_1 b_{\text{Cu}^*\text{Cu}}b_{\text{NiCu}}N_{i\text{Ni}1} +\Phi_{\text{Cu}}.
\end{equation}
Using the above expression the Cu spectrum can be decomposed into, or constructed from, components associated with Cu atoms with each value of 
$N_{i\text{Ni}1}$, which allows us to equate changes in the Cu spectrum to changes in the system's `$N_{i\text{Ni}1}$ histogram' for Cu.

This is done in Fig. \ref{fig:figure_Cu5Ni5}(b) for [Ni$_5$/Cu$_5$], where we have used $\Gamma_{\text{life}}=0.3$eV in the simulated spectra.
At $\sigma=0$ Cu sites exhibit only two possible environments, $N_{i\text{Ni}1}=0$ or 4, with the former corresponding to the 3 `central' monolayers of the 
5 monolayer Cu stack and the latter corresponding to the 2 `edge' monolayers. On the whole Cu sites exhibit higher values of $N_{i\text{Ni}1}$ as Cu diffuses 
into the Ni region, leading to a shift in the spectrum to low binding energies as $\sigma$ increases. The width of the spectrum increases as the 
$N_{i\text{Ni}1}$ histogram becomes `flat' near $\sigma=1.5$, and then narrows again as $\sigma\to\infty$ and the histogram tends to that corresponding 
to the random alloy Ni$_{0.5}$Cu$_{0.5}$. Note that, as we found earlier to be the case for the $B/A_T/B$ systems, the disorder broadening is not maximized at 
the random alloy configuration. An explanation for this is as follows. The $N_{i\text{Ni}1}$ histograms for random alloys -
substitutionally disordered systems with homogeneous concentration profiles - are binomial distributions. For inhomogeneous concentration profiles, such 
as those for [Ni$_5$/Cu$_5$] at $\sigma\neq\infty$, the histograms are not constrained to be binomial distributions - they are free to be `flatter'. 
Therefore systems with inhomogeneous concentration profiles can exhibit significantly larger disorder broadenings than random alloys - as is
borne out in Figs. \ref{fig:figure_Cu5Ni5}(b), \ref{fig:fwhm_BAB} and \ref{fig:trajectories_CuNi}.

\begin{figure}
\centering
\includegraphics[width=0.5\textwidth]{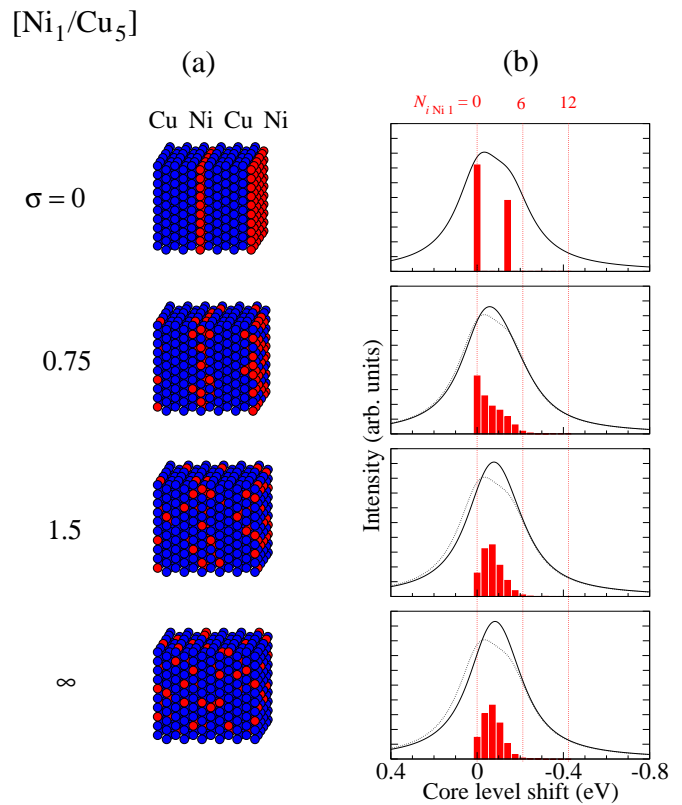}
\caption
{The details of this figure are the same as Fig. \ref{fig:figure_Cu5Ni5}, except that this figure pertains to the alloy system 
[Ni$_1$/Cu$_5$].}
\label{fig:figure_Cu5Ni1}
\end{figure}

As another example, consider [Ni$_1$/Cu$_5$], which we earlier pointed out exhibits a disorder narrowing. Fig. \ref{fig:figure_Cu5Ni1} is
analogous to Fig. \ref{fig:figure_Cu5Ni5}, but for [Ni$_1$/Cu$_5$]. Again, we used $\Gamma_{\text{life}}=0.3$eV in the simulated spectrum. 
At $\sigma=0$ the Cu sites in the 2 
edge monolayers of a 5 monolayer Cu stack have $N_{i\text{Ni}1}=4$, while the Cu sites in the 3 central monolayers have $N_{i\text{Ni}1}=0$. This
is in fact exactly the same situation as for [Ni$_5$/Cu$_5$]. As $\sigma$ is increased the Cu sites begin to exhibit other values of 
$N_{i\text{Ni}1}$; some Cu sites begin to `occupy' the $N_{i\text{Ni}1}=1-3$ `states' which were unoccupied at $\sigma=0$. The same also occurs
in [Ni$_5$/Cu$_5$]. However, comparing Figs. \ref{fig:figure_Cu5Ni5} and \ref{fig:figure_Cu5Ni1} it can be seen that there is a significant 
occupation of the $N_{i\text{Ni}1}>4$ states in [Ni$_5$/Cu$_5$] which does not occur in [Ni$_1$/Cu$_5$] on account of the lack of Ni.
Hence in [Ni$_1$/Cu$_5$], loosely speaking, the Cu sites spill from their $N_{i\text{Ni}1}=0$ and $N_{i\text{Ni}1}=4$ states into only the
$N_{i\text{Ni}1}=1-3$ states as $\sigma$ is increased, which results in a narrowing of the spectrum in this system.

\subsubsection{Comparison with experimental results}
Experimental Cu spectra for [Ni$_5$/Cu$_5$] and [Ni$_5$/Cu$_2$] at various temperatures were obtained in Refs. \onlinecite{Holmstrom_2006,Granroth_2009}.
However, the spectra were found to depend strongly on the material used to cap the sample, and also on the photon energy used \cite{Granroth_2009}. This
is due to the finite escape depth of the photoelectrons, which leads to Cu sites near the surface being `over-represented' in the Cu 
spectra. Our model results do not account for such surface effects; they pertain to the deep bulk. Hence a quantitative comparison between our model 
results and most of the experimental spectra is meaningless. The possible exception is the spectra for [Ni$_5$/Cu$_5$] in Ref. \onlinecite{Granroth_2009} 
taken using a photon energy of 6030eV - which corresponds to a relatively low surface sensitivity. These spectra reveal a shift of $\approx-0.2$eV in 
[Ni$_5$/Cu$_5$] upon heating over the temperature range corresponding to the transition from $\sigma=0$ to $\sigma=\infty$. This is in excellent agreement 
with our results: as can be seen from Fig. \ref{fig:trajectories_CuNi}, from $\sigma=0$ to $\sigma=\infty$ the model predicts a shift in the Cu CLS of 
$\approx-0.2$eV for [Ni$_5$/Cu$_5$].
In the future it would be interesting to perform model calculations which take into account surface effects, and which 
therefore can be compared directly with the results of Refs. \onlinecite{Holmstrom_2006,Granroth_2009}.

\subsection{Ternary systems}
Finally, we consider ternary multilayer systems - consisting of three species $A$, $B$ and $C$. 
Recall that the $X$ CLS distribution reflects the $X$ charge distribution (Eqn. 
\eqref{DeltaEBk}), which itself is determined by charge transfer between $X$ and non-$X$ sites (Eqn. \eqref{Qi}). In binary systems Eqn. \eqref{Qi} gives
\begin{equation}
\begin{split}
Q_i=&\Lambda b_{BA}\sum_{\beta=1}^{\infty}g_{\beta}N_{iB\beta}\quad\text{if $i\in A$} \\
Q_i=&-\Lambda b_{BA}\sum_{\beta=1}^{\infty}g_{\beta}N_{iA\beta}\quad\text{if $i\in B$},
\end{split}
\end{equation}
where we have used the fact that $b_{YX}=-b_{XY}$.
Without loss of generality let $b_{BA}>0$, which corresponds to choosing species $A$ to be the most electronegative of $A$ and $B$.
From the above equations it can be seen that the electropositivity difference $b_{BA}$ between species $B$ and $A$ acts only as a scale factor for
the $A$ and $B$ charge distributions; altering $b_{BA}$ has no effect on the \emph{qualitative} nature of the $A$ and $B$ charge distributions.
Ternary systems are more complicated. The analogous equation to the above for species $A$ in a ternary system is
\begin{equation}
Q_i=\Lambda\biggl[ b_{BA}\sum_{\beta=1}^{\infty}g_{\beta}N_{iB\beta} + b_{CA}\sum_{\beta=1}^{\infty}g_{\beta}N_{iC\beta} \biggr].
\end{equation}
Note that altering $b_{BA}$ affects the charge transfer between site $i$ and local $B$ sites, while leaving the transfer with local $C$ sites unaffected; and
conversely if $b_{CA}$ is altered. Hence altering $b_{CA}$ or $b_{BA}$ non-trivially alters the $A$ charge distribution. 
In this sense ternary systems are extremely rich.
An exhaustive survey of what can be expected from such systems is beyond the scope of this work. We instead limit ourselves to two ternary systems which 
we have found to exhibit somewhat counterintuitive behaviors. Furthermore, we limit our discussion to the qualitative aspects of these systems, with
a focus on understanding these behaviors. For both systems species $A$ is the focus of our attention.

\subsubsection{Periodic ternary multilayer: $[B_5/A_5/C_5]$}
\begin{figure}
\centering
\includegraphics[width=0.5\textwidth]{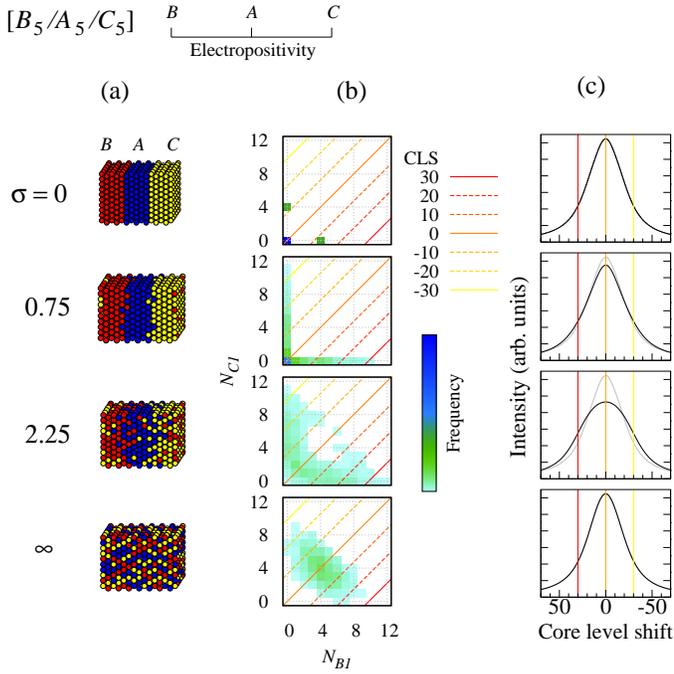}
\caption
{(Color online) Schematic illustration of [$B_5$/$A_5$/$C_5$] at various $\sigma$ (a); and the corresponding
$(N_{B1},N_{C1})$ histograms and environment vs. CLS maps for species $A$ (b), and model spectra (c). In (b), the green-blue squares represents
the frequency of $A$ sites with the environment $(N_{B1},N_{C1})$, and yellow-red curves are contours of constant CLS determined according to
Eqn. \eqref{DeltaEBi_nn_ter}. In (c) the dotted curves are spectra for $\sigma=0$, and the location of some of the CLS 
contours in (b) are also shown.}
\label{fig:figure_B5A5C5}
\end{figure}

The first ternary system we consider is $[B_5/A_5/C_5]$. Again, we use Eqn. \eqref{conc_sigma} to model the interface
roughening; Fig. \ref{fig:figure_B5A5C5}(a) gives a schematic illustration of the system at the $\sigma$ we considered.
For our calculations we set the electropositivities of each species to be $b_B=-1$, $b_A=0$ and $b_C=1$: the electropositivity of species $A$
is exactly halfway between those of species $B$ and $C$, with species $B$ being the most electronegative and species $C$ being the most electropositive.
With regards to the other model parameters we set $\Phi_A=0$, $aR_{\text{WS}}=2.0$, $R_{\text{WS}}=1$, $b_{A^*A}=1$, and $\Gamma_{\text{life}}=35$.
Fig. \ref{fig:figure_B5A5C5}(c) shows the simulated $A$ spectra at each $\sigma$.
As is evident from the figure, the mean of the $A$ CLS distribution is independent of $\sigma$. 
Hence here one cannot use the mean to characterize $\sigma$ in this system - one must use the FWHM. This stems from the choice of species 
electropositivities. Charge transfer from $B$ to $A$ and $C$ to $A$ is always equal and opposite on account of the electropositivity of $A$ being exactly 
between that of $B$ and $C$. This, in conjunction with the symmetry of the system, always yields a global mean $A$ charge of 0 regardless of $\sigma$,
and hence, from Eqn. \eqref{DeltaEBk}, a $\sigma$-independent mean CLS. 
Another interesting feature of this system is that the $\sigma=0$ and $\sigma=\infty$ spectra are ostensibly 
indistinguishable. Hence unambiguously determining $\sigma$ in this system from the $A$ spectrum alone is difficult.

The method described earlier for rationalizing spectral changes in binary systems can be extended to ternary systems. The analogous equation to 
Eqn. \eqref{DeltaEBi_nn_bin} for species $A$ in a ternary system is
\begin{equation}\label{DeltaEBi_nn_ter}
\Delta E^{\text{B}}_i\approx -a\Lambda g_1 b_{A^*A}\Bigl[b_{BA}N_{iB1}+b_{CA}N_{iC1}\Bigr] +\Phi_A.
\end{equation}
Note that here $\Delta E^{\text{B}}_i$ depends on the environment of $i$ through the pair $(N_{iB1},N_{iC1})$; there is a mapping from $(N_{iB1},N_{iC1})$ to
$\Delta E^{\text{B}}_i$. Hence the $A$ spectrum reflects the `$(N_{iB1},N_{iC1})$ histogram' of species $A$. This is illustrated in Fig.
\ref{fig:figure_B5A5C5}(b-c).
Fig. \ref{fig:figure_B5A5C5}(b) illustrates the frequency of $A$ sites with each $(N_{iB1},N_{iC1})$ (i.e., the $(N_{iB1},N_{iC1})$ histogram for species
$A$), as well as the CLS for each $(N_{iB1},N_{iC1})$ - contours of constant CLS in `$(N_{iB1},N_{iC1})$-space' are drawn.
If an environment $(N_{iB1},N_{iC1})$ is exhibited by a high frequency of sites, then there is a spike in the CLS spectrum at the corresponding 
CLS. This allows rationalization of the evolution of the $A$ spectrum with $\sigma$. At $\sigma=0$ there are only 3 possible environments for $A$ sites: 
$(N_{iB1},N_{iC1})=(4,0)$, which corresponds to a site adjacent to the $B$ region; $(0,4)$, which corresponds to a site adjacent to the $C$ region; and 
$(0,0)$, which corresponds to a site in the center of the $A$ region - surrounded by $A$ sites. At $\sigma=0.75$ there is a small amount of intermixing
at the interfaces. This leads to $A$ sites near the $B$ region exhibiting environments $(1,0)$, $(2,0)$, $(3,0)$, etc.; and $A$ sites near the $C$ region
exhibiting environments $(0,1)$, $(0,2)$, $(0,3)$, etc. Note that at this point no $A$ sites have both $B$ \emph{and} $C$ nearest neighbors, and hence
the frequency of any environment $(N_{iB1},N_{iC1})$ is only non-zero if $(N_{iB1},N_{iC1})=(x,0)$ or $(0,x)$ for any $x$. 
With regards to the spectrum, the `spreading out' of the histogram along
the left and bottom edges means that environments are exhibited which correspond to more extreme CLSs, i.e., $\Delta E^{\text{B}}_i\approx 30$ and 
$\approx-30$. Hence the spectrum broadens from $\sigma=0$ to 0.75. 
At $\sigma=2.25$ the interface roughening is large enough that there are $A$ sites with both $B$ and $C$
neighbors, and hence there is a non-zero frequency for environments away from the left and lower edges of the $(N_{iB1},N_{iC1})$ histogram.
At $\sigma=\infty$ we have a ternary random alloy, and the $(N_{iB1},N_{iC1})$ histogram corresponds to that of a trinomial distribution.

\subsubsection{Intermixing near an immiscible thin film: $A_5/B_5/C$}
\begin{figure}
\centering
\includegraphics[width=0.5\textwidth]{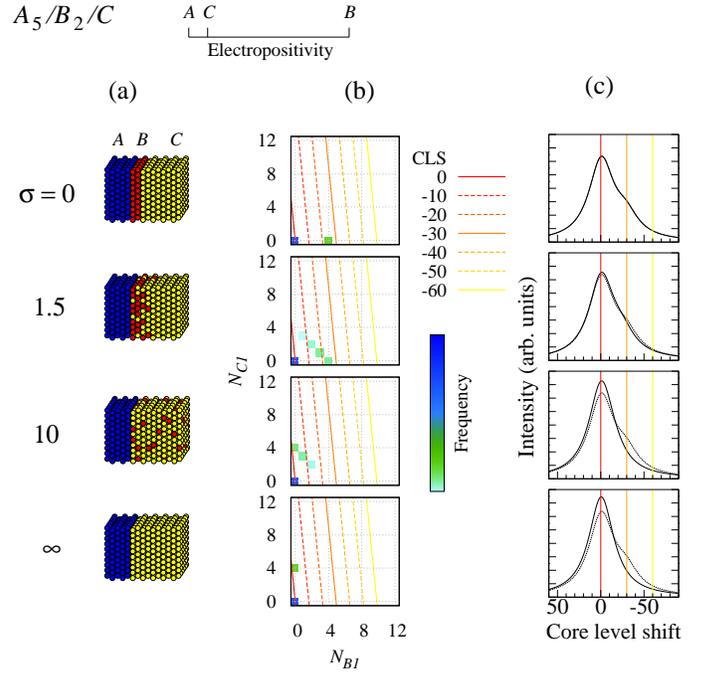}
\caption
{The details of this figure are the same as Fig. \ref{fig:figure_B5A5C5}, except that this figure pertains to the alloy system 
$A_5$/$B_2$/$C$}
\label{fig:figure_A5B2C}
\end{figure}

The final system we consider is $A_5/B_5/C$, with $A$ constrained to be immiscible in $B$ and $C$, and the electropositivities set to $b_A=0$, 
$b_C=0.1$, and $b_B=1$. The electropositivities correspond to the following situation: species $A$ is the most electronegative; species $B$ is the most
electropositive; and species $C$ has an intermediate electropositivity very close to that of species $A$.
This system is interesting because, like [Ni$_1$/Cu$_5$] discussed earlier, it exhibits a disorder narrowing.
Constraining $A$ to be immiscible in $B$ and $C$ renders Eqn. \eqref{conc_sigma} unsuitable. We therefore instead modeled the interface roughening for
this system using the equation
\begin{equation}\label{conc_sigma_wall}
c^l_X(\sigma)=\sum_{m=1}^{\infty}\Bigl[\gamma^{l-m}(\sigma)-\gamma^{l+m}(\sigma)\Bigr]c^l_X(\sigma=0)
\end{equation}
for $X=B,C$, where we have used the convention that the interface between species $A$ and $B$ at $\sigma=0$ is located between monolayers 0 and 1.
The resulting concentration profiles are illustrated in Fig. \ref{fig:figure_A5B2C}(a).
Similarly to Fig. \ref{fig:figure_B5A5C5}, Figs. \ref{fig:figure_A5B2C}(b) and \ref{fig:figure_A5B2C}(c) show the $(N_{iB1},N_{iC1})$ histogram 
and simulated spectra for species $A$ at each of the considered $\sigma$. 
For this system we used the same $\Phi_A$, $aR_{\text{WS}}$, $R_{\text{WS}}$, $b_{A^*A}$ and $\Gamma_{\text{life}}$ as for $[B_5/A_5/C_5]$.

Earlier we saw that [Ni$_1$/Cu$_5$] exhibits a disorder narrowing. In general, given that the $A$ spectrum of a system reflects the $A$ charge distribution
- as follows from Eqn. \eqref{DeltaEBk} -  a disorder narrowing for $A$ 
occurs when the introduction of substitutional disorder `quenches' the width of the $A$ charge 
distribution. This occurs here. Species $B$ and $A$ transfer a certain amount of charge, which results
in an $A$ charge distribution with a certain width at $\sigma=0$. As $\sigma$ is increased, more $C$ sites come within the charge-transfer range of
the $A$ region. Given the tiny electropositivity difference between species $A$ and $C$, there is almost no charge transfer between $A$ and $C$ sites. 
Hence the influx of $C$ sites to the $A$ interface acts to reduce the charges of the edge $A$ sites, bringing them closer to that of
the `non-edge' $A$ sites. This corresponds to a reduction in the width of the $A$ charge distribution, and hence also the core level spectrum.
Alternatively, one can explain the disorder narrowing in terms of the evolution of the $(N_{iB1},N_{iC1})$ histogram for species $A$ 
(Fig. \ref{fig:figure_A5B2C}(b)). At $\sigma=0$ the environment of the $A$ sites on the edge monolayer is $(N_{iB1},N_{iC1})=(4,0)$, which steadily
transitions to $(N_{iB1},N_{iC1})=(0,4)$ as $\sigma\to\infty$. The latter environment has a CLS closer to that of the non-edge $A$ sites, i.e., 
$(N_{iB1},N_{iC1})=(0,0)$, and hence the spectrum narrows as $\sigma$ is increased.

\section{Summary and discussion}\label{sec:summary}
Above we have presented a model for core level shifts in alloys, and have used it to add insight into the relationship between atomic 
environment, charge transfer and disorder broadening in complex systems. Our key result is that the mapping between the distribution of atomic
environments and core level spectra is often  counterintuitive. For instance, systems with inhomogeneous concentration profiles can exhibit
significantly larger disorder broadenings than is possible in random alloys, and even a `disorder narrowing' in some cases. For the correct 
interpretation of experimental spectra for complex systems, it is crucially important to understand such phenomena.

The model can be easily adapted to treat other core level spectroscopies, the most prominent of which is Auger electron spectroscopy. 
Hence we expect that it should find widespread use as a framework in which to interpret experimental results. However, it is by no means a
panacea. Recall that the model is underpinned by the NRA-CEFM - which itself is a particular case of the CEFM. Implicit in the NRA-CEFM are a number of
approximations which may be problematic. One is the \emph{spherical approximation} - that only the monopole moments of the charge distribution in each 
site are considered for the purposes of evaluating the Madelung energy. One must go beyond this approximation in order to obtain a quantitatively 
accurate description of the electron density within disordered alloys \cite{Ruban_2002_2,Pinski_1998}.
This is especially true for regions near surfaces. A generalization of the CEFM has been described in Ref. \onlinecite{Bruno_2008} which does not rely 
upon the spherical approximation. While the NRA-CEFM could be generalized in an analogous manner, it is not clear whether this would be fruitful. The 
strength of the NRA-CEFM over the `general' CEFM is its simplicity, with which comes a small loss in accuracy relative to the CEFM. It is not clear
 whether the gain in accuracy achieved by generalizing the NRA-CEFM to go beyond the spherical approximation is worth the resulting loss in simplicity.
Another potentially problematic approximation implicit in the NRA-CEFM is that the nuclei of the system under consideration form an undistorted 
crystal lattice. The breakdown of this approximation can have far-reaching consequences. The addition of distortions to the crystal lattice of CuAu results 
in a reversal of the average relationship between a site's CLS and its number of unlike nearest neighbors \cite{Marten_2009}. The reasons for this 
are not known, and warrant further investigation. The NRA-CEFM, suitably modified to treat lattice distortions, may add insight into this phenomenon, 
though it would be optimistic to expect that anything more than a qualitative understanding could be achieved.

It should be borne in mind that the aforementioned approximations, while implicit in the NRA-CEFM, are also utilized in many \emph{ab initio}
calculations, and are not expected to preclude the model from making \emph{at least} qualitatively accurate predictions. A more problematic limitation 
of the model is that its free parameters are not known \emph{a priori}: they must be obtained from \emph{ab initio} calculations or by other means. 
Fortunately these parameters are highly transferable between systems; for details see Refs. \onlinecite{Bruno_2008,Underwood_2013}. In the future
we intend to calculate these parameters for a wide range of alloys. This would enable the model to be readily applied to many systems.

\begin{acknowledgements}
This work was supported by the Engineering and Physical Sciences Research Council.
\end{acknowledgements}

\end{document}